\documentclass[review]{elsarticle}

\usepackage{amsmath}    
\usepackage{amsfonts}  
\usepackage{amssymb}
\usepackage{graphicx}   

\usepackage{dcolumn}
\usepackage{bm}

\usepackage{natbib}

\usepackage{mathptmx}      
\usepackage{latexsym}
\newcommand{\be}{\begin{equation}}
\newcommand{\ee}{\end{equation}}
\newcommand{\bea}{\begin{eqnarray}}
\newcommand{\eea}{\end{eqnarray}}

\newcommand{\bbm}{\begin{bmatrix}}
\newcommand{\ebm}{\end{bmatrix}}

\newcommand{\iGAj}{{j}}

\newcommand{\cliffconj}[1] { \bar{#1} }

\begin{document}

\begin{frontmatter}

\title{A brief study of time}

\author{James M.~Chappell}
\address{School of Electrical and Electronic Engineering, University of Adelaide, SA 5005, Australia}
\cortext[mycorrespondingauthor]{Corresponding author}
\ead{james.chappell@adelaide.edu.au}

\author{John G.~Hartnett }
\address{Institute for Photonics \& Advanced Sensing (IPAS), and the School of Physical Sciences, University of Adelaide, SA 5005 Australia}

\author{Azhar Iqbal}
\address{School of Electrical and Electronic Engineering, University of Adelaide, SA 5005 Australia}

\author{Nicolangelo Iannella}
\address{Institute for Telecommunications Research, The University of South Australia, Mawson Lakes SA 5095}

\author{Derek Abbott}
\address{School of Electrical and Electronic Engineering, University of Adelaide, SA 5005, Australia}

\begin{abstract}
Understanding the nature of time remains a key unsolved problem in science.  Newton in the Principia asserted an absolute universal time that {\it `flows equably'}. Hamilton then proposed a mathematical unification of space and time within the framework of the quaternions that ultimately lead to the famous Minkowski formulation in 1908 using four-vectors.  The Minkowski framework is found to provide a versatile formalism for describing the relationship between space and time in accordance with relativistic principles, but nevertheless fails to provide deeper insights into the physical origin of time and its properties. 
In this paper we begin with a recognition of the fundamental role played by three-dimensional space in physics that we model using the Clifford algebra multivector.  From this geometrical foundation we are then able to identify a plausible origin for our concept of time.  This geometrical perspective also allows us to make a key topological distinction between time and space, with time being a point-like quantity. The multivector then allows a generalized unification of time and space within a Minkowski-like description.  
\end{abstract}

\begin{keyword}
\texttt{Time \sep Geometric algebra \sep Quaternions \sep Minkowski \sep Spacetime}
\MSC[2010] 00-02 \sep  83-02
\end{keyword}

\end{frontmatter}


\section{Introduction}
\label{intro}

Historically, there have been many attempts to understand the nature of time and provide a rigorous mathematical definition. 

One of the most influential ideas regarding time was developed by Sir Isaac Newton in 1686 described in {\it Principia} 
Book 1, \emph{``Absolute, true, and mathematical time, of itself, and from its own nature, flows equably without relation to anything external, \dots'' }~\cite{Newton1686}  This idea of the steady flow of time was overturned by relativity theory, which showed that the apparent rate of clocks actually depends on relative motion as well as the relative strength of the gravitational field, between source and observer. As a way to salvage Newton's definition we can limit its application to the observer in the rest frame of the particle under observation. This definition of Newton, that includes the idea of the flow of time, may appear to also imply an arrow of time, however Newton's laws are all in fact time reversible. Hence Newton appears to be describing some sort of non-directional flow associated with time.  So in the rest frame of the particle, exactly what sort of clock could we be referring to?  One obvious candidate would be the de~Broglie frequency $ \nu $ that is assumed to be associated with every particle of energy $ E $, which in its own rest frame is given by $ \nu = E/h = m c^2/h $, where $ m $ is the mass of the particle and $ h $ is the Planck constant.  The de~Broglie frequency can be associated with the {\it Zitterbewegung} frequency predicted by Schr{\"o}dinger in 1930 and now confirmed experimentally~\cite{Roos2010}. This clock based on the steady de~Broglie frequency would appear to satisfy Newton's definition of time.

We can also attempt to draw further insight from the representation of time in the Minkowski spacetime framework $ [ t, x, y, z] $, where $ t $ is the time coordinate.  This appears to imply that time is a physical linear dimension, orthogonal to the three other space dimensions and so naturally leads to the concept of time being the fourth dimension of experience.  This description of a local time given by the Minkowski four-vector implies that time now effectively exists in the infinite past and infinite future, and leads to the idea of the block universe. A caveat on the idea of treating time as a fourth Euclidean-type dimension is that it does not appear possible, at least in the macroscopic world, to freely travel in the time dimension as is possible with space dimensions. Hence, these difficulties with the Minkowski formulation leads us to look elsewhere for a more fundamental description of time.

Regarding the understanding of time drawn from the principles of quantum mechanics, Feynman states, {\it ``Sum over histories indicates the direction of our ordinary clock time is simply a path in space which is more probable than the more exotic directions time might have taken otherwise.''}~\cite{Feynman1965}  This statement forces us to acknowledge that perhaps our intuitive ideas about the flow of time garnered from observing the macroscopic world are in reality only statistical in nature.

\subsection{The origin of our concept of time}
\label{origintime}

As a way to identify the most fundamental conceptions of time we now look at the first origins of our ideas about time from observations of the natural world.
Perhaps the most striking phenomena from which we form the concept of time is the rotation of the earth on its axis producing the regular day-night cycle.  Historically this has been dominant in mankind's measurement of time. Thus the earth itself becomes the clock that can be used to measure the duration between events in units of days. Indeed, in ancient times, the time period of one day was also commonly subdivided into fractions through the use of sundials. To measure a 15 degree change in the position of the sun, meaning 1 hour, using a sundial was relatively straightforward. It would also be natural to compare its rate with some other cycle in order to confirm its regularity, and the next most obvious cyclic phenomena would be the lunar cycle, which has the advantage of providing a clock during night hours.  By extending this reasoning to the other visible planets, the whole solar system then becomes a giant clock mapped onto the zodiac of the `fixed' stars.  This then naturally leads to the idea of an external regular rate of some time within which events can be placed.  As stated by Newton regarding the distinction between measured time and some assumed universal time {\it `Relative, apparent, and common time, is some sensible and external (whether accurate or inequable) measure of duration by the means of motion, which is commonly used instead of true time; such as an hour, a day, a month, a year.'}~\cite{Newton1686}  That is, time as measured by various clocks is only considered an approximation by Newton to some universal absolute time.  This idea appeared reasonable at the time as the sun was considered the center of the universe about which the planets orbited.  The nature of the milky way galaxy as a rotating gravitationally bound object as well as extra-galactic objects had not yet been established.
Today of course atomic clocks have been developed that can produce extremely regular vibrations that are stable over very long periods and so can replace the solar system clock.  

So far in our discussion we have now identified three different types of clocks that are based on different principles of nature.  The rotation of the earth on its axis based on the principle of the conservation of angular momentum, the lunar orbital period based on the law of gravity and atomic clocks based on the laws of quantum mechanics. Nevertheless these three clocks have been measured to run at essentially the same rate and so this self consistency reinforces the idea of an external dimension of time. It is indeed presumed that this correspondence is exact provided we allow for variations in the solar system orbits caused by losses due to tidal fraction etc. 
If the principles of special and general relativity are also utilized, then we can adapt clock rates measured on the surface of the earth to other frames of reference, including those subject to gravity and thus attempt to produce a consistent definition of time throughout the universe.  However, due to the relativity of simultaneity, while we can allocate a time sequence of events from the perspective of the earth frame we cannot establish a single set of clock times for events common to all observers.  

Since the work of Hubble, an alternative definition of time has become possible, namely that related to the expansion of the Universe.  This is not an inertial, gravitational or atomic clock and indeed, due to the proposed era of rapid cosmic inflation, perhaps not even a steady one. Nevertheless using a general relativistic model, such as the Friedmann-Lema{\^\i}tre-Robertson-Walker metric we can define a cosmic time for the universe of approximately 13.7 billion years.  This general relativistic definition of time is also different to other definitions of time due to the recently discovered accelerating expansion of the universe, which indicates, if correct, that the universe apparently will never reverse direction, and so appearing to indicate a direction or arrow to this definition of time.
This one-way behavior is also consistent with the law of entropy, that postulates the inexorable path of all systems to maximum disorder. The Second Law applies to all isolated systems and has been extensively tested, but it is an unknown if it applies to the Universe as a whole. After all, we do not have an ensemble of universes to test the hypothesis on.  The observation of the expanding universe perhaps also providing a way out of the Loschmidt paradox, that states that time irreversible physical laws cannot be produced from time reversible laws.
The observed fact, that the universe is currently not in state of maximum disorder, appears to indicate that the universe is currently far from equilibrium in a low entropy thermodynamic state.  Recently the principle of entropy and the flow of time has been underpinned with an explanation based on the flow of quantum entanglement.~\cite{Popescu2009,Wooters1983} This theory proposes that the process of an object coming to thermal equilibrium, for example, is actually a process of an object becoming increasingly correlated or entangled with the environment.
 While indeed appearing to provide an explanation for entropy and the flow of time based on entanglement, Popescu nevertheless accepts that the nature of time itself remains {\it ``one of the greatest unknowns in physics''}~\cite{Popescu2009}. Indeed the resolution of the concept of time also considered one of the key issues in quantum gravity. Regarding the present-day definition of time, Smolin states {\it `I believe there is something basic we are all missing, some wrong assumption we are all making....My guess is that it involves two things: the
foundations of quantum mechanics and the nature of time.}~\cite{Smolin2006}
It should also be noted that the weak force also appears to distinguish between the two time directions. This indicates that the quantum level may indeed also provide some direction to time.
Nevertheless, at the microscopic level, time generally appears to freely move in both directions, as shown for example, in elementary particle interactions. One conception of the positron is that it is an electron moving backward in time and Marjorana particles are their own anti-particle, and so simultaneously move forwards and backwards in time.  
One further type of clock commonly used in archeology is that based on the decay of radioactive isotopes.  This is a non-periodic, statistical, directional clock commonly used to determine the age of matter.

In summary, we noted that all clocks appear to tick at relatively consistent rate, however the stability and the mutual agreement of all these various measures of time fundamentally rely on the stability of the laws of nature.  Hence we are still left with the question of what is the most fundamental definition of time.  Einstein, utilized exclusively the light clock as a basis for developing special relativity and so appeared to consider this type of clock as the more fundamental.  Indeed the de-Broglie frequency based on the {\it Zitterbewegung} is essentially a light clock. We noted the reversible type of clocks based on cyclic or rotational motion as well as a time direction given perhaps by the nature of the expanding universe and entropy.  Overall, the existence of time appears to be based on various types of regular motion, whether cyclic or a steady flow. If time is intrinsically based on movement, then this would then indicate that we need physical space in order for the concept of time to arise and so time therefore arises with the origin of the spatial universe.  This leads us to briefly review some philosophical and religious views of time.

\subsection{Religious views of time}
\label{religiontime}

The idea of the creation of time concomitant with the origin of the universe is actually in agreement with several philosophical and religious conceptions of time. For example, Augustine's statement, which has also been reiterated more recently by Stephen Hawking, asserts that {\it ``The world was made, not in time, but simultaneously with time''}~\cite{Augustine1972}.  
Indeed this view of time by Augustine could be seen to be inspired by the Torah that states that also describes the beginning of time {\it ``And there was evening and there was morning, the first day.''}{\it King James Version}  Hinduism also has a similar view of time as a continuously repeating cyclic process.  
However, nearly all religious traditions also refer to time as not only having a beginning but also having an end, with time appearing sandwiched between two eras of timelessness. Hinduism making the claim that time only exists while we are in a state of becoming or change that is followed by an era of unity and timelessness.
This concept of timelessness outside spacetime is indeed consistent with the Minkowski spacetime structure.  This is because time and space are only defined within the spacetime continuum itself, and therefore outside this structure time is essentially undefined and so timeless\cite{Yourgrau2005}.

\subsection{Psychological aspects of time}
\label{psychology}

It might appear irrelevant to consider psychological factors in considering a fundamental physical definition of time.
For example, human perception intrinsically utilizes the concept of past, present and future, upon which Einstein commented {\it ``...for us physicists believe the separation between past, present, and future is only an illusion, although a convincing one.''}\footnote{Einstein, Albert, Letter to Michele Besso's Family. Ref. Bernstein, Jeremy., A Critic at Large: Besso. The New Yorker (1989)}  The relativity of simultaneity derived by Einstein indeed rules out a universal `now' throughout the universe as presumed by Newton with his formulation of an absolute time. The relativity of simultaneity in turn follows from the Einstein simultaneity convention (ESC), that is based on the synchronization of two clocks separated in space by the sending and receiving of light signals.  The mathematical framework for describing the concept of past and future for a local observer in relativity theory then becomes the light cone.  

Thus the psychological aspects become a complicating factor in our analysis of time. This may have something to do with the fact that the brain and visual processing apparatus of the observer are continually faced with the problem of needing to process a seemingly infinite panorama of experience within a finite brain of limited processing power.  Some of the tricks used by the brain in order to reduce the processing load for temporal information perhaps include conceptual frameworks such as the past, present and future, the steady linear `flow' of time and the arrow of time.  Also the concept of cause and effect, can lead to the concept of an arrow of time in situations where there is an increase in entropy.
These mostly human perceptions of time have resulted in various ontological theories of time being developed such as Presentism, Eternalism or the A-theory and the B-theory of time~\cite{oaklander2004ontology}. We consider these theories however as mostly reflecting human psychology.  For example, the concept of the past can be interpreted as a property of memory and the concept of the future a property of the facility of imagination. Hence, we now seek to rather underpin these general theories of time through identifying the true physical nature of space and time as reflected in elementary natural laws.

\subsection{Algebraic formulations of time}
\label{algebratime}

One of the current roadblocks to progress in order to mathematically describe space and time appears to be that despite over a hundred years of intense development there is still no generally accepted algebraic description of three-dimensional physical space.
This fact is quite surprising in that, one of the specific goals of nineteenth century science was to find this algebra~\cite{Crowe1967}.  The objective was initially led by Hamilton who produced the quaternion algebra through generalizing the two-dimensional complex numbers to three dimensions. 	Due to the success of complex numbers in algebraically describing the properties of the plane Hamilton believed that quaternions would therefore fully describe the algebra of three-dimensional space. This then lead to the first attempt at a rigorous mathematical definition of a unified space and time. Hamilton wrote the quaternion as
\be
q = t + x_1 \boldsymbol{i} + x_2 \boldsymbol{j} + x_3 \boldsymbol{k} ,
\ee
where $ t, x_1, x_2, x_3 \in \Re $ and the three basis vectors are subject to the well known quaternionic relations $ \boldsymbol{i}^2 = \boldsymbol{j}^2 = \boldsymbol{k}^2 = \boldsymbol{i} \boldsymbol{j} \boldsymbol{k} =  -1 $. 
Hamilton's quaternions form a four-dimensional associative normed division algebra over the real numbers represented by $ {\mathbb{H}}  $.  
Hamilton defined $ \boldsymbol{x} = x_1 \boldsymbol{i} + x_2 \boldsymbol{j} + x_3 \boldsymbol{k}  $ as a vector quaternion to take the role of Cartesian vectors in order to describe three-dimensional space, but then also proposed, nearly 50 years before Minkowski, that if the scalar `$t$' was identified with time then the quaternion can be used as a representation for a unified four-dimensional spacetime.  Hamilton stating
{\it ``Time is said to have only one dimension, and space to have three dimensions. ... The mathematical quaternion partakes of both these elements; in technical language it may be said to be `time plus space', or `space plus time': and in this sense it has, or at least involves a reference to, four dimensions.'}~\cite{Graves1882}
Indeed squaring the quaternion we find
\be
q^2 = t^2 - \boldsymbol{x}^2 + 2 t \boldsymbol{x}
\ee
the scalar component $ t^2 - \boldsymbol{x}^2 $ thus producing the invariant spacetime distance, as required by Einstein.
This definition of time within a quaternionic spacetime thus has the key property of producing the invariant distance and indeed provides a viable definition of space-time.  However Hamilton was not the first to conceive of a four-dimensional space-time structure, as Lagrange much earlier around 1797 asserts  {\it Mechanics may be regarded as a four-dimensional geometry, and mechanical 
analysis as an extension of geometrical analysis.}

Following the work by Hamilton, Minkowski after indeed considering the quaternions as a possible description of spacetime, chose rather to extend the Gibbs three-vector system with the addition of a time coordinate to create a four-component vector producing the modern description of spacetime, as a four-vector
\be
X = [t, \boldsymbol{x} ] .
\ee
Defining an involution $ \cliffconj{X} =  [t, -\boldsymbol{x} ] $ we then produce the invariant distance
\be
X \cdot \cliffconj{X} =  [t, \boldsymbol{x} ] \cdot  [t, -\boldsymbol{x} ] = t^2 - \boldsymbol{x}^2 ,
\ee
as required. 

Comparing these two descriptions we can see that they provide some subtle differences.  To begin with, Hamilton's description views time as a part of the description of three-dimensional space, indeed this was the motivation for developing the quaternion algebra in the first place.  
On the other hand, with the Minkowski formulation, the immediate implication is that time is an additional Euclidean-type dimension.  Although this assumption is ameliorated by the fact that time contributes an opposite sign to the metric distance and so it is quite distinct from a regular four-dimensional Cartesian vector.

It is an historical fact that the vector quaternions were found difficult to work with and not suitable to describe Cartesian vectors and were replaced by the Gibbs vector system in use today.  The reason for Hamilton's failed attempt to algebraically describe three-dimensional space, is that by generalizing the complex numbers he only produced the rotational algebra for three dimensions.  That is, because of the isomorphism in the plane of U(1) $\cong$ SO(2) he only generalized the rotational algebra to U(2), and thus had not given a full generalization of the plane to three dimensions that included Cartesian vectors. 
If we wish to include a Cartesian vector then the generalization of the plane to three dimensions becomes $ C\ell(\Re^3 ) $, an eight-dimensional Clifford algebra and so twice the size of the quaternion algebra. Now, as we found, the quaternion algebra provides a natural description of space and time, it is therefore logical to consider how the Clifford generalization adds to this understanding.

\section{Clifford multivector spacetime}
\label{clifford}

A Clifford geometric algebra $ C\ell \left(\Re^n\right) $ defines an associative real algebra over $ n $ dimensions and in three dimensions $ C\ell \left(\Re^3\right) $ is eight-dimensional. 
In this case we can adopt the three quantities $ e_1, e_2, e_3 $ for basis vectors that are defined to anticommute in the same way as Hamilton's quaternions, but unlike the quaternions these quantities square to positive one, that is $ e_1^2 = e_2^2 = e_3^2 = 1 $.  
Also, similar to quaternions, we can combine scalars and the various algebraic components into a single number that is now called a multivector
\be \label{multivectorFull}
X = t + x_1 e_1 + x_2 e_2 + x_3 e_3 + n_1 e_2 e_3 + n_2 e_3 e_1 + n_3 e_1 e_2 + b e_1 e_2 e_3 ,
\ee
where $ t, x_1, x_2, x_3, n_1, n_2, n_3, b \in \Re $.
Now defining $ j = e_1 e_2 e_3 $ we find the dual relations $ j e_1 = e_2 e_3 $, $ j e_2 = e_3 e_1 $ and $ j e_3 = e_1 e_2 $, which allows us to write
\be \label{multivectorWithVectors}
X = t + \boldsymbol{x} + j \boldsymbol{n} + j b ,
\ee
with the vectors $ \boldsymbol{x} = x_1 e_1 + x_2 e_2 + x_3 e_3 $ and $ \boldsymbol{n} = n_1 e_1 + n_2 e_2 + n_3 e_3 $.

It can be shown that quaternions are isomorphic to the even subalgebra of the multivector, with the mapping $ {\boldsymbol{i}} \leftrightarrow e_2 e_3 $, $ {\boldsymbol{j}} \leftrightarrow e_1 e_3 $, $ {\boldsymbol{k}} \leftrightarrow e_1 e_2 $ and the Gibbs vector can be replaced by the vector component of the multivector.  
Thus within Clifford's unified system we can absorb the quaternion as $ q = t + j \boldsymbol{n}  $ and a Gibbs vector $ \boldsymbol{x} $.

We define Clifford conjugation on a multivector $ X $ as
\be
\cliffconj{X} = t - \boldsymbol{x} - j \boldsymbol{n} + j b .
\ee
We define the {\bfseries amplitude} of a multivector $ X $ as $ | X |^2 = X \cliffconj{X}  $ that gives
\be \label{AmplitudeSquared}
| X |^2 = t^2 - \boldsymbol{x}^2 + \boldsymbol{n}^2 - b^2 + 2 \iGAj \left ( t b - \boldsymbol{x} \cdot \boldsymbol{n} \right )
\ee
forming a commuting ``complex-like''~\footnote{We refer to this as a complex-like number because the trivector $ \iGAj $ is commuting and squares to minus one and all other quantities are real scalars.} number. 

Clifford conjugation that produces the multivector amplitude turns out to be the only viable definition for the metric as it is the only option that produces a commuting resultant and is thus an element of the center of the algebra.  This is essential as we require the metric distances to be isotropic in order to be consistent with the principles of relativity.  Once again we can observe the required invariant distance $ t^2 - \boldsymbol{x}^2 $ appearing in the metric.  Thus the multivector description of spacetime has an immediate advantage of leading naturally to the invariant distance whereas it is axiomatically assumed in the Minkowski formulation.

An important point to note for the multivector, is that in order to produce a meaningful metric, which consists of a combination of its various elements, then all of these components must be measured in the same units, as shown in Eq.~(\ref{AmplitudeSquared}).  Now, beginning from 1983, the General Conference on Weights and Measures (CGPM) decided that the speed of light should be assumed constant and that a meter was then the distance traveled by light in a specified time interval equal to $ 1/c $ seconds.  The CGPM defines the speed of light as a universal constant $ c = 299,792,458 $m/s. Hence both time and distance are now measured in units of seconds, and it is therefore natural to adopt these units for all components of the multivector.  Distances typically measured in meters therefore appear in the multivector with the conversion $ \boldsymbol{x} \rightarrow \boldsymbol{x}/c $ and so are in units of seconds.  Interestingly, from the perspective of the multivector, $ c $ is simply taking the role of a units conversion factor, and so therefore the value of $ c $ is obviously invariant between observers.  Hence Einstein's apparently surprising assertion of the invariance of the speed of light in 1905, in fact follows directly from the nature of space and time described by the multivector.   Hence one confusion regarding time could arise due to the poor selection of units, as in order to properly relate space and time they should be measured in the same units, as now done by CGPM. 

The multivector generalization of the quaternions now allows a full algebraic description of three-dimensional space, as the scalar, vector, bivectors and trivectors components now correspond directly with the geometrical quantities of points, lines, areas and volumes found in three dimensions. Additionally these four quantities describe the range of physical quantities described as scalars, vectors, pseudovectors (or axial vectors) and pseudoscalars.
Now, similar to quaternions, as we also identify time as the scalar part of space, time is now imputed the geometrical meaning and topology of a scalar point-like quantity.  This can be contrasted with the linear description of space as vectors and so we can see now a sharp geometrical distinction between space and time when described within the multivector.  
Time is often considered as being a very ineffable concept and not something that is tangible like space.
That is, while the space dimensions appear readily available to our senses the time dimension appears strangely invisible. This fact can now perhaps be explained by the fact that time being a scalar or point-like quantity is therefore essentially invisible to the senses.

\section{Discussion}
\label{discuss}

Minkowski showed that the concept of a global space or a global time on their own do not exist but only that a unified spacetime exists.  This concept is now extended with an eight-dimensional spacetime multivector that can also be defined globally, and the individual components, such as time or space, can only be defined locally.  Of course this effectively means that the spacetime structure itself is outside time, because time is now only defined within it.  Indeed general relativity effectively incorporates this idea with a dynamical spacetime structure used to define gravity\cite{Yourgrau2005}. 
Also, consistent with this viewpoint, tests on quantum systems have demonstrated that time is an emergent phenomenon for internal observers but absent for external observers~\cite{Moreva2014}.
Specifically for the spacetime structure described by the multivector, in Eq.~(\ref{multivectorWithVectors}), we typically select the  scalar time parameter as the evolution parameter, finding
\be \label{multivectorWithVectors}
\left ( 1 + \frac{d \boldsymbol{x}}{d t} + j \frac{d \boldsymbol{n}}{d t} + j \frac{d b}{d t} \right ) d t.
\ee
This enables the concept of velocity $ \frac{d \boldsymbol{x}}{d t} $ to be produced, for example.
However, nearly as conveniently we could have selected the trivector term $ b $ representing torsion as the evolution parameter.  Hence conventional time, represented here as the scalar, now loses it primacy as the evolution parameter.

In order to aid our intuition, we can use a rubber sheet analogy to give a geometrical picture of the multivector and hence a more tangible understanding of space and time. The scalar component $ t $ can represent the expansion or contraction of the sheet at each point, the displacement $  \boldsymbol{x} $ the linear distortion within the sheet, the bivector $ \iGAj \boldsymbol{n} $ the planar twist and the trivector $  \iGAj b $ the three-dimensional torsion of the sheet. It is interesting that the scalar can refer to the expansion of the rubber sheet, especially as we have already found an association of time, which we represented as a scalar, and cosmic time measured with the expansion of the universe.  This identification of time with expansion is also consistent with Newton's concept of the non-directional flow of time if we assume that space is expanding uniformly in all directions. 

The Clifford multivector provides a generalized description of spacetime and significantly views time and space as simply two particular geometrical properties that we abstract from three-dimensional physical space.  It is also natural to  ask then regarding the other two geometrical components of the multivector, that of bivectors and trivectors. These of course we refer to commonly as area and volume but also refer to physical quantities having the attributes of pseudovectors and pseudoscalars, respectively.  We propose that these ignored quantities in the Minkowski spacetime description help explain our confused understanding regarding the nature of time as their contribution to the metric is generally labeled as simply a property of time rather than distinguishing an additional time-like component to physical space.
Now, inspecting the metric in Eq.~(\ref{AmplitudeSquared}) we can see that the bivector component also provides the correct signature for the time component as well as the scalar component that we initially selected, giving the modified metric for the spacetime interval of $ (dt^2 + d\boldsymbol{n}^2) - d\boldsymbol{x}^2 $, ignoring for the present discussion the other contributions to the metric.  This appears to indicate that the whole quaternion $ T = t + \iGAj \boldsymbol{n}  $ could therefore be used to represent the property of time. This makes sense as it then unifies the two key aspects of time that we have identified that of the reversible rotational part $ \iGAj \boldsymbol{n} $ and the irreversible expanding part as the scalar $ t $.  If this line of thinking is adopted then time becomes a four-dimensional quaternionic quantity, and indeed other multi-time theories of spacetime have already been investigated.~\cite{tifft1996three,bars2001survey,Cole1980comments,dorling1970dimensionality,Muses1985}  
Hence it is now apparent that we distinguish a second type of time-like quantity that we here refer to as rotational time represented by the bivector $ \iGAj \boldsymbol{n} $ in distinction to scalar expansional time given by the scalar $ t $.

This distinction can now perhaps explain the conundrum of why it is possible to move freely in space but not in time. If we have a coffee cup, for example, then clearly we can move it in a certain direction and then reverse this translation by sliding it back again to its starting point. For rotating the cup it appears that we can do the same, however, if we consider time as related to the microscopic rotations at the atomic level and all the spin axes are randomly aligned then it is not possible to reverse all these microscopic spin directions simultaneously. This is in distinction to the space direction due to the molecular bonding of the atoms where it is indeed possible to reverse the spatial movement of each atom simultaneously. 
When we move to the level of fundamental particles the rotational nature of time allows fundamental particles to move backwards in time if they invert their rotation, as in the CPT symmetry.  
Hawking has also recently introduced a new theory in an attempt to explain the origin of the universe that incorporates the idea of imaginary time in the no-boundary proposal~\cite{Hawking1983}.  This could also perhaps be included as part of the bivector rotational time term $ j \boldsymbol{n} $.
The third and final geometrical component of physical space is the trivector $ j b  $ that describes the torsion of physical space that we could identify as a third helical form of time, nicely represented by a circularly polarized photon.  It should be noted though that this term has a space-like contribution to the metric.

\section{Conclusion}
\label{conclusion}

We began from the assumption that the eight-dimensional Clifford geometric algebra $ C\ell(\Re^3) $ provides the correct algebraic description of three-dimensional physical space, as supported by recent research.  The multivector naturally describing algebraically the four geometric elements of space, that of points, lines, areas and volumes, as shown in Eq.~(\ref{multivectorWithVectors}).  These four quantities also describe the physical quantities referred to as scalar, vectors, pseudovectors and pseudoscalars and found to be a natural language to describe physical theories in three dimensions~\cite{Chappell2014IEEE,chappell2011revisiting}.
We thus consider that the eight-dimensional Clifford multivector describing three-dimensional space is the basis upon which we abstract such local concepts as time and space.  

When we represent spacetime within the Clifford multivector, in order to be consistent with Minkowski spacetime we find that time needs to identified with the scalar component and space with the vector component of the multivector.  This thus gives a view of time as the geometric point-like quantities of space.  With vectors describing space, we thus now have a geometrical union of time and space as the points and lines, respectively.  Thus the correct topology of time can thus be proposed as a point-like entity that is in distinction from the Minkowski formulation that implies a linear topology. 

Also, using the rubber sheet analogy we noted that the scalar can represent an expansion of the sheet and so nicely correlated with cosmic time as defined by an expanding universe.  This idea of time represented by expansion also neatly ties in with the Newtonian idea of the non-directional flow of time.  Thus the scalar nature of time can be seen as arising from counting cyclic phenomena such as the de~Broglie-{\it Zitterbewegung} cycles or alternatively as the general expansion of space. 

So `what is time?' We conclude that it is simply one of the properties that we abstract from the geometry of the three-dimensional space, specifically the geometrical point-like quantities.  

It was noted that the bivectors were also time-like and so could be included as a description of time.  Thus time can become quaternionic combining scalar and bivector components.  Thus this appears to provide a way to unify different ideas of time, where the scalar attribute can represent the irreversible expansional aspects of time and the bivector the reversible rotational attributes of time, respectively. The trivector torsional aspect of space also provides a possible third aspect to our description of time.  

Ultimately the full union of time and space needs to include all the geometric elements of point-like, linear, areal and volume elements and so implies that a four-vector is insufficient, but that we require an eight-dimensional spacetime event multivector, incorporating the scalar, linear, rotational and the torsional aspects of space, as shown in Eq.~(\ref{multivectorWithVectors}). Science, currently ignoring these additional components within spacetime could help explain the confusion in our current understanding of time.
It is suggested therefore that the Clifford multivector provides a suitable foundation for describing the full nature of the spacetime continuum within science.

\section*{References}









\bibliographystyle{elsarticle-num}

\bibliography{quantum}

\end{document}